\documentclass[a4paper]{article}
\usepackage{tcolorbox}[2015/01/01]
\usepackage{amsfonts}
\usepackage[round]{natbib}
\usepackage{amssymb}
\usepackage{amsthm}
\usepackage{latexsym,amsmath}
\usepackage{graphicx}
\usepackage{afterpage}
\usepackage{listings}
\usepackage[final]{pdfpages}
\usepackage{lineno}
\usepackage{relsize}
\usepackage{nameref}
\usepackage{verbatim}
\usepackage{float}
\usepackage{multirow}
\usepackage{algorithm}
\usepackage{colortbl}
\usepackage[noend]{algpseudocode}
\usepackage{cleveref}
\usepackage{mathrsfs}
\usepackage{caption}
\usepackage{subcaption}
\usepackage{setspace}
\usepackage[parfill]{parskip}

\newcommand{\PP}{\mathbb{P}}

\def\tr(#1){{\rm trace}(#1)}
\def\Exp(#1){{\mathbb E}(#1)}
\def\Exps(#1){{\mathbb E}\sparen(#1)}

\begin{document}
\newtheorem{theorem}{Theorem}[section]
\newtheorem{lemma}[theorem]{Lemma}
\newtheorem{proposition}[theorem]{Proposition}
\newtheorem{corollary}[theorem]{Corollary}
\newtheorem{conjecture}[theorem]{Conjecture}
\newtheorem{definition}[theorem]{Definition}
\newtheorem{example}[theorem]{Example}
\newtheorem{remark}[theorem]{Remark}
\newtheorem{question}[theorem]{Question}
\newtheorem{notation}[theorem]{Notation}
\numberwithin{equation}{section}

%\linenumbers

\section*{How many sites?  Methods to assist design decisions when collecting multivariate data in ecology}

\textbf{Ben Maslen$^1$, Gordana Popovic$^{1,2}$, Adriana Verg\'es$^{2,3,4}$, Ezequiel Marzinelli$^{4,5,6}$, David Warton$^{1,2}$}\

\small{\textit{$^1$School of Mathematics and Statistics, University of New South Wales, NSW 2052, Australia}}

\small{\textit{$^2$Evolution and Ecology Research Centre, University of New South Wales, Sydney, NSW 2052, Australia}}

\small{\textit{$^3$School of Biological, Earth, and Environmental Sciences, University of New South Wales, Sydney, NSW 2052, Australia}}

\small{\textit{$^4$Sydney Institute of Marine Science, 19 Chowder Bay Rd, Mosman NSW 2088, Australia}}

\small{\textit{$^5$The University of Sydney, School of Life and Environmental Sciences, Coastal and Marine Ecosystems, Sydney NSW 2006, Australia}}

\small{\textit{$^6$Singapore Centre for Environmental Life Sciences Engineering, Nanyang Technological University, Singapore}}

%\textbf{Contact details (respectively):}\\
%b.maslen@unsw.edu.au, g.popovic@unsw.edu.au, a.verges@unsw.edu.au, e.marzinelli@sydney.edu.au, david.warton@unsw.edu.au

\textbf{Corresponding author:}\\
Ben Maslen\\ 
89 Dawson St, Cooks Hill NSW 2300\\
b.maslen@unsw.edu.au

\textbf{Short running headline:}\\
Tools for planning multivariate studies

%\textbf{Data accessibility statement:} Should the manuscript be accepted, the data will be archived in the Dryad public repository and the data DOI will be included at the end of the article.

%\textbf{Number of:}

%Abstract words: 149

%Main text words: 4075

%Text box words (respectively): 33, 15

%References: 35

%Figures: 4 (+2 in Appendix)

%Tables: 0

%Text boxes: 2

\pagebreak

\begin{abstract}

\noindent \begin{enumerate}
    \item Sample size estimation through power analysis is a fundamental tool in planning an ecological study, yet there are currently no well-established procedures for when multivariate abundances are to be collected. A power analysis procedure would need to address three challenges: designing a parsimonious simulation model that captures key community data properties; measuring effect size in a realistic yet interpretable fashion; and ensuring computational feasibility when simulation is used both for power estimation and significance testing.
    \item Here we propose a power analysis procedure that addresses these three challenges by: using for simulation a Gaussian copula model with factor analytical structure, fitted to pilot data; assuming a common effect size across all taxa, but applied in different directions according to expert opinion (to ``increaser", ``decreaser" or ``no effect" taxa); using a critical value approach to estimate power, which reduces computation time by a factor of 500 with little loss of accuracy.
    %\item Our specified approach to estimating power for multivariate abundances is shown to meet these challenges. Our simplified approach to effect size specification takes a complicated multivariate power estimate into a feasible, yet informed procedure, whilst our Gaussian copula model simulates data with similar properties to the pilot data that it has been tuned through. Our critical value approach is also shown to be an effective power estimate and reduces computation time from days to minutes, making the procedure practically applicable for researchers.
    \item The procedure is demonstrated on pilot data from fish assemblages in a restoration study, where it was found that the planned study design would only be capable of detecting relatively large effects (change in abundance by a factor of 1.5 or more).
    \item The methods outlined in this paper are available in accompanying \texttt{R} software (the \texttt{ecopower} package), which allows researchers with pilot data to answer a wide range of design questions to assist them in planning their studies.
\end{enumerate}

\end{abstract}

\textbf{Keywords:}
copula, multivariate, power analysis, restoration, sample size, simulation, software, statistics, study design

\pagebreak

\section*{Introduction}

In planning any study it is important to consider how large an effect (``effect size") can be detected by the intended study design \citep{ rosenthal1994parametric, fritz2007required,kelley2012effect}. Such information is often obtained via a power analysis of preliminary data, which can help design a study that has a good chance of detecting effects of ecological interest \citep{cohen1992statistical,cohen2013statistical}. While texts for ecologists frequently discuss the importance of power analysis, and related techniques, in sample size determination \citep{gerrodette1987power,johnson2015power,green2016simr}, there is little guidance for the ecologist on how to undertake such an analysis for multivariate data, i.e.\ multiple response variables measured per sample. This is particularly the case if analysing abundance or presence-absence data simultaneously collected for many different taxa; hereafter (\emph{multivariate abundances}).\\

There are three key challenges that would need to be addressed in developing a power analysis procedure for multivariate abundance data. \emph{Challenge 1} is to design a simulation model to randomly generate realistic multivariate abundance data, reflecting key properties of the data to be collected. It should be possible to use pilot data, when available, to tune the settings of the simulation model, to generate data that ``looks like" the pilot data. Some simulation approaches are available in the literature \citep[e.g.][]{xu2010industrial}, however with limited ability to tune using pilot data. Other methods have also been developed using a Bray-Curtis distance based approach \citep{irvine2011power} or by considering the abundances as continuous random variables and calculating Euclidean distances \citep{collins2000method, angeler2009statistical}. These approaches however ignore important mean-variance assumptions and therefore subsequent analyses can be misleading \citep{warton2012distance}. Multivariate modelling approaches have been developed recently that could be used to address this, including Gaussian copulas \citep{popovic2018general,anderson2019pathway} and hierarchical models \citep{warton2015model, ovaskainen2017make}. \emph{Challenge 2} is measuring effect size. When many taxa are to be sampled, many parameters need to be specified \textit{a priori} that will capture the size and nature of the effect the study has been designed to detect. Decisions about these parameters are to be made based on relatively little information, and need to be captured by a simple, interpretable effect size measure in order for results to be useful for study design \citep{kelley2012effect}. \emph{Challenge 3} is making the power analysis procedure computationally efficient. Hypothesis testing procedures for multivariate abundance data typically use resampling \citep{anderson2001new,wang2012mvabund}, conventionally recalculating a test statistic at least 1000 times across resampled datasets. A power analysis would require this to be done for each of say 1000 simulated datasets, such that estimating power for a typical multivariate abundance dataset would take hours or days.\\

This paper proposes a power analysis procedure for multivariate abundance data that addresses each of the above three challenges. \emph{Challenge 1} is addressed using a Gaussian copula-factor analysis model to simulate data in a parsimonious fashion, an approach for which methods to tune the model using training data were only recently proposed \citep{popovic2018general}. \emph{Challenge 2} is addressed using a simple parameterisation for effect size that requires \textit{a priori} information only concerning taxa that are likely to be affected, and the direction of the effect. \emph{Challenge 3} is addressed using a novel ``critical value" approach to power analysis of resampled statistics, which can reduce computation time from days to minutes. The procedure will be illustrated on a marine habitat restoration project which involves regular monitoring of ecological communities, where we are interested in the number of samples required for a future monitoring period in order to likely detect community differences in abundance across treatments, as well as the size of effects that are able to be detected under different sampling designs.

\section*{Materials and Methods}

\subsection*{Operation Crayweed Restoration Project}
\begin{figure}[H]
						\centering
						\begin{subfigure}{.32\textwidth}
							\centering
							\includegraphics[width=.90\linewidth,height=35.5mm]{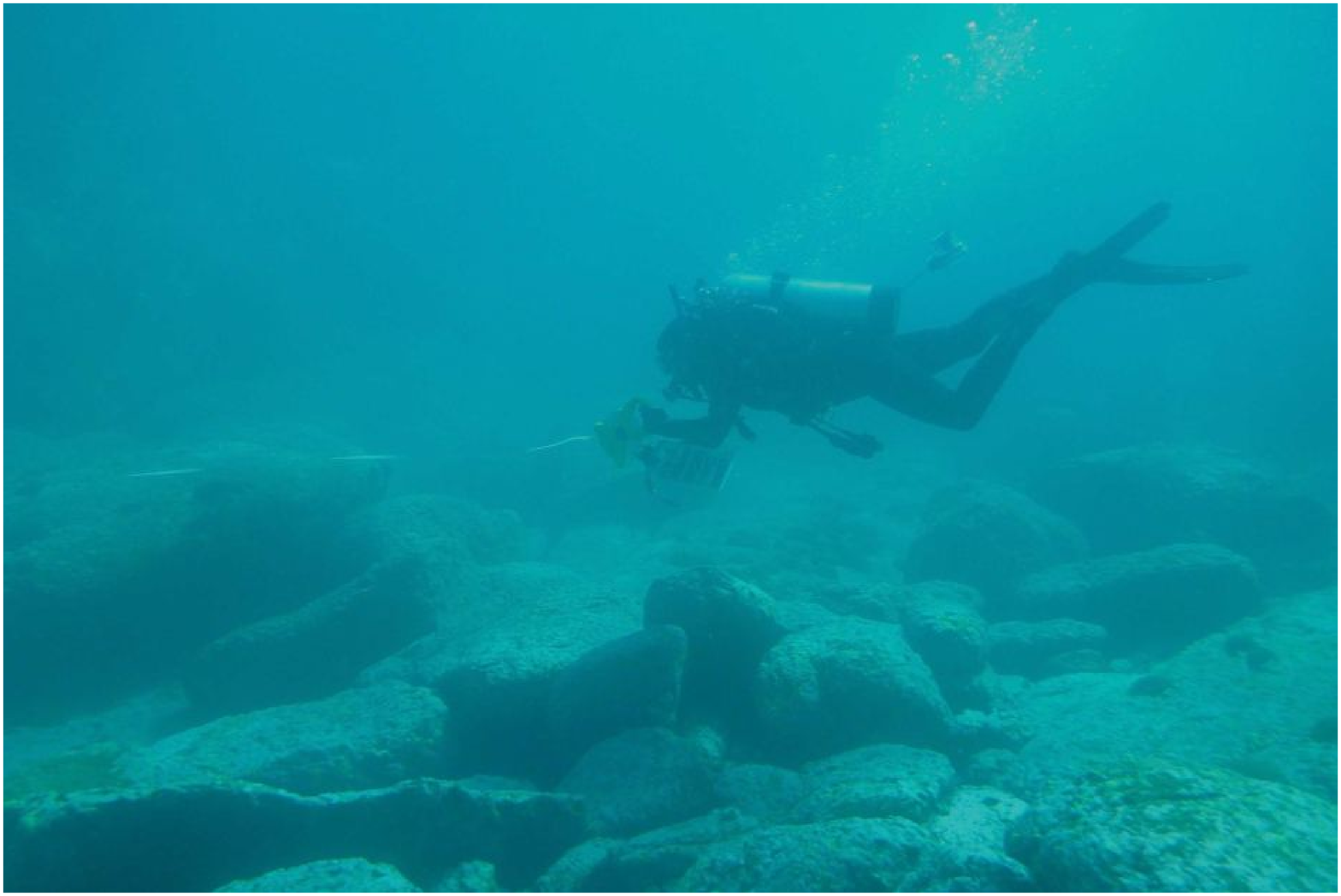}
							\caption{Control}
							\label{fig:sub1}
						\end{subfigure}%
						\begin{subfigure}{.3\textwidth}
							\centering
							\includegraphics[width=.95\linewidth]{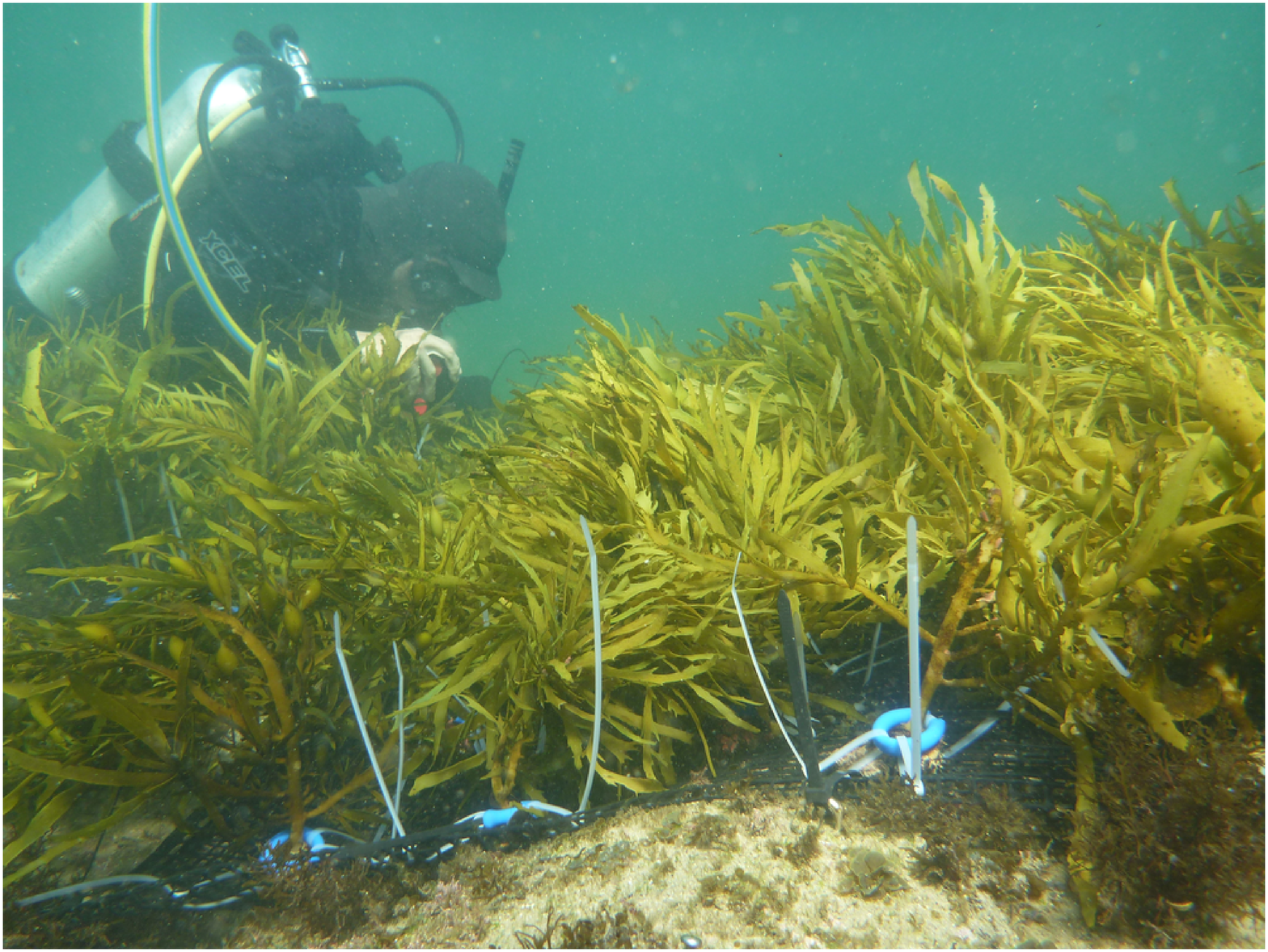}
							\caption{Restored}
							\label{fig:sub2}
						\end{subfigure}
						\begin{subfigure}{.3\textwidth}
							\centering
							\includegraphics[width=.95\linewidth]{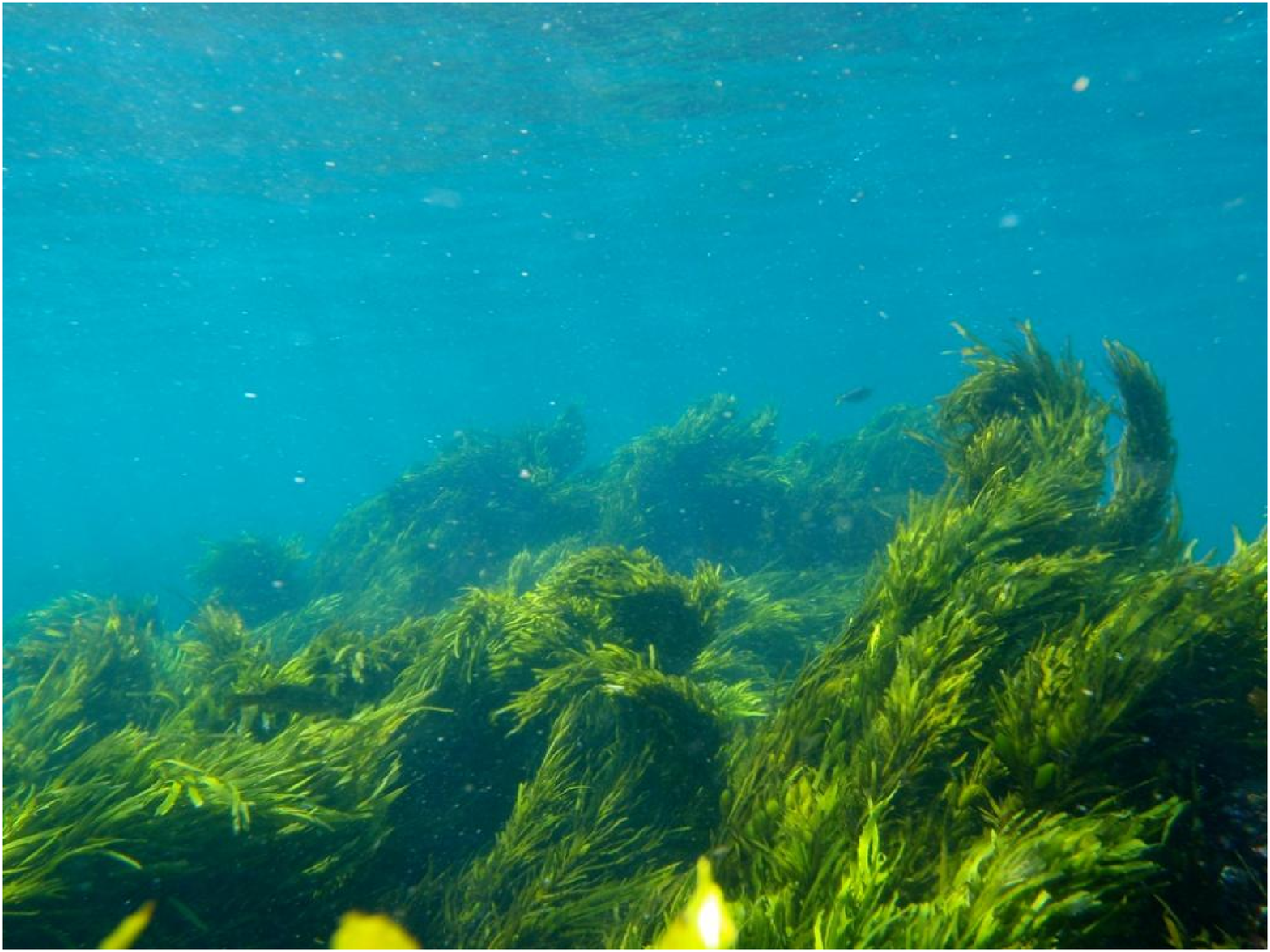}
							\caption{Reference}
							\label{fig:sub3}
						\end{subfigure}
						\caption{\small{Images of characterised environments from each treatment level. Control sites are characterised as urchin barren habitats in Sydney shallow rocky reefs where crayweed has been lost. Restored sites are similar to control sites where crayweed has recently been transplanted. Reference sites are shallow rocky reefs outside of the Sydney region with extant crayweed populations.}}
							\label{fig:treatments}
					\end{figure}

Researchers within the Operation Crayweed Restoration Project in Sydney are restoring the locally extinct macro-algae \textit{Phyllospora comosa} \citep[``crayweed": see][]{coleman2008absence, campbell2014towards, verges2020operation} and are interested in the effect of this restoration on associated ecological communities \citep{marzinelli2014restoring, marzinelli2016does, wood2019restoring}. Pilot data have already been collected, where the abundance of fish species in nine open ocean sites have been recorded. We are interested in observing if there is a change in mean fish abundance between three treatments: control (sites in Sydney without crayweed), restored (similar to control sites where crayweed has recently been planted) and reference (sites north and south of Sydney with extant crayweed forests), as in Figure \ref{fig:treatments}. There are plans to collect more data in the future, however there is an upper bound of approximately 24 possible spatially independent restored or control coastal bays/sites within the Sydney region and surroundings.

%FIGURE 1

We are interested in answering the following experimental design questions from the pilot data, which will act as the motivating examples for this paper:
\begin{enumerate}
    \item How many sites are required in order to likely detect $20\%$ differences between treatments?
    \item Under the maximum balanced sampling design of 12 sites for each treatment, what are the size of effects that are likely to be detected?
\end{enumerate}

\subsection*{Power analysis}
Consider a situation where the objective of the study is to test a specific null hypothesis ($H_0$) by looking for evidence of some alternative ($H_1$). The effectiveness of any given study design and testing procedure can be evaluated using power, i.e. the probability of rejecting the null hypothesis $H_0$ given that a particular alternative hypothesis $H_1$ is true. Power generally increases as sample size and effect size increases, and variability decreases \citep{cohen2013statistical}. Thus, given an understanding of the variability to be expected in the data, power can be used to work out how large a sample size is needed to detect an effect of a given size, or what sizes of effects can be detected in study of a given sample size \citep{cohen1992statistical,cohen2013statistical}.\\

\begin{figure}[H]
					\centering
			\includegraphics[scale=0.45]{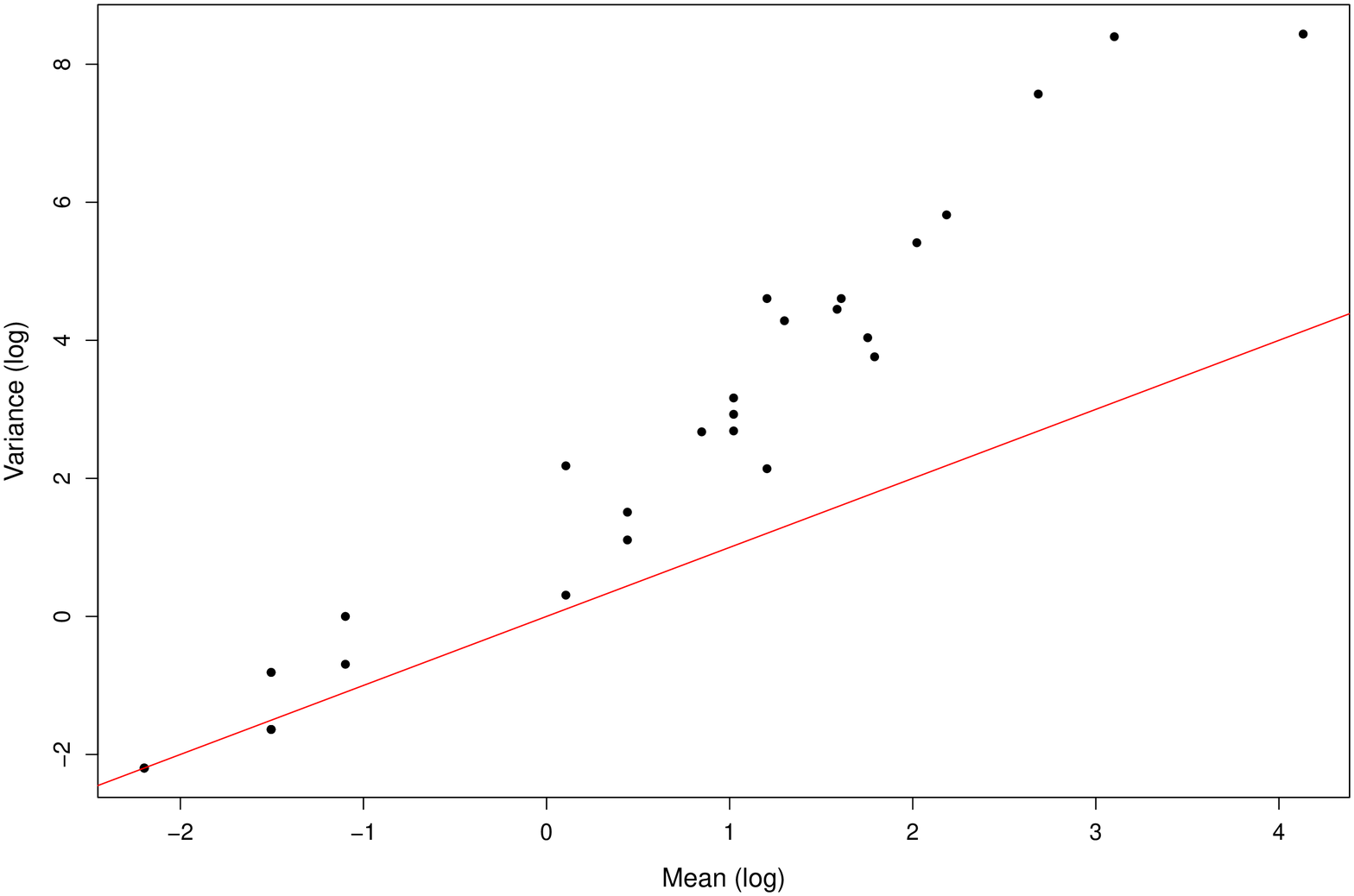}
			\caption{\small{Mean-variance plot of fish abundances for p=34 species. The red line depicts the
mean-variance assumption under a Poisson model (mean=variance). The
variance appears larger than the mean for species with larger abundances, indicating
over-dispersion relative to the Poisson distribution, with the negative binomial distribution being preferred for this dataset.}}
			\label{fig:meanvar}
	\end{figure}

\begin{figure}[H]
					\centering
			\includegraphics[scale=0.5]{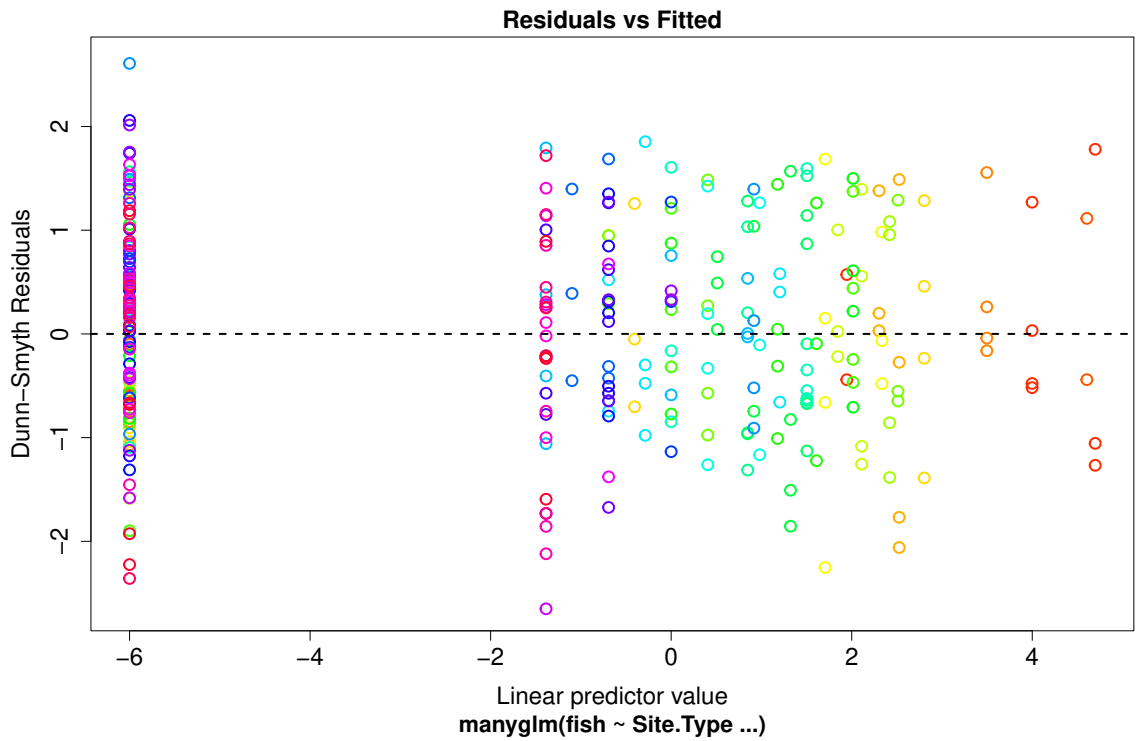}
			\caption{\small{Diagnostic plot of a negative binomial model fitted to the fish abundance data from the Crayweed Restoration Project using \texttt{manyglm}. The large cluster of data on the left hand side of this graph refers to observations predicted to have zero abundance in a treatment. The lack of fan shape in the residual vs. fitted plot implies the negative binomial distribution has adequately accounted for the mean-variance relationship in the data shown in Figure \ref{fig:meanvar}.}}
			\label{fig:diag}
	\end{figure}

In order to undertake a power analysis, it is necessary to work through the following steps:
\begin{enumerate}
\item Specify a model for data, which captures key properties of the data that are expected to be collected. For multivariate abundances, as in Operation Crayweed, this is a non-trivial task (\emph{Challenge 1}).
\item Decide on a measure of effect size that is ecologically meaningful. When there are many taxa, as for fish assemblages sampled in Operation Crayweed, there are many different effect size parameters that need to be considered (\emph{Challenge 2}).
\item Decide on the testing procedure. Hypothesis tests of multivariate abundance data typically use resampling to ensure valid inference \citep{anderson2001new,wang2012mvabund}; in this paper testing will make use of a generalised estimating equations approach \citep[using the \texttt{mvabund} package]{wang2012mvabund} that is increasingly common in ecology. For the crayweed fish abundance data, we assumed abundances have a negative binomial distribution, with diagnostic plots suggesting this adequately captured the mean-variance trend in the data (Figures \ref{fig:meanvar} and \ref{fig:diag} in Supporting Information).
\item Estimate power. In some simpler settings this can be done analytically \citep{cohen2013statistical}, however for multivariate abundances this needs to be done by simulation, generating data under the assumed model, then applying the testing procedure for each simulated dataset, and recording the proportion of times the null hypothesis was rejected. When using a resampling-based testing procedure, this involves two levels of simulation and will be very computationally intensive (\emph{Challenge 3}).
\end{enumerate}
The following sections detail the proposed solutions to the three challenges identified above.

\subsection*{Challenge 1 -- Data generating model}
\label{ref:Simulatingapproach}

A data generating model is needed that can capture key properties of multivariate abundance data that will be collected. Multivariate abundances are discrete, with many zeros, and high dimensional, with a large number of responses $p$ relative to the sample size $n$. The ability to specify a parametric statistical model that can be fitted to multivariate abundances is a relatively recent development \citep{popovic2018general, warton2015so}. This is an important advance because it allows the simulation model to be tuned to pilot data in order to generate data statistically similar to what will actually be observed when the study is undertaken. In particular, power can be strongly affected by mean abundance, variability, and correlation across taxa \citep{warton2011regularized}, all of which can vary considerably from one study to another. These data properties need to be tuned to the study in question for a power analysis to be informative.\\

This paper adopts a Gaussian copula approach. To date, copulas have rarely been used in ecology \citep{popovic2018general, anderson2019pathway, popovic2019untangling}. A specific advantage of a copula approach is that it specifies a marginal (unconditional) model, making parameters more interpretable. For example, if we set an effect size parameter that ensures a two-fold change in mean abundance between treatment and control groups, we can be sure that there will in fact be a two-fold change. In hierarchical models this does not always happen, because they operate as conditional models, which can induce some surprising behaviour when interpreted marginally \citep{breslow1995bias,lin1996bias,gurka2011avoiding}. A nice feature of the copula model is that it assumes the same marginal model as the testing procedure to be used here \citep{wang2012mvabund}, and so is a suitable simulation model when using a Generalised Estimating Equation (GEE) procedure. Copulas are also used in \cite{anderson2019pathway}, although to compare different test procedures, rather than for sample size determination.\\

A discrete Gaussion copula \citep{popovic2018general,popovic2019untangling} can model correlated discrete data $y_{ij}$ using latent Gaussian variables $z_{ij}$:

\begin{equation}\tag{1}
    Y_{ij}={F}_j^{-1}(\Phi(Z_{ij}))
\end{equation}

Where $F_j$ is the marginal distribution for taxon $j$ (e.g. Poisson with mean $\lambda = \exp (XB)$). $\Phi$ is  a multivariate Gaussian distribution with zero mean and covariance structure $\Sigma$,

\[
z_{ij} \sim \mathcal{N}_p(0,\Sigma).
\]

This model is estimated using maximum likelihood to obtain $\hat{F}_j$ and $\hat{\Sigma}$. To simulate new multivariate abundances we simply simulate new latent variables $z_{ij} \sim \mathcal{N}_p(0,\hat{\Sigma})$ and then obtain abundances by transforming $y_{ij}= {F^*}_j^{-1}(\Phi(z_{ij}))$, where $\hat{F}^*_j$ has been parameterised via an effect of interest.\\

One difficulty with this approach however is estimating a covariance structure across responses. With $p$ taxa, there are  $p(p-1)/2$ pairwise correlations to estimate, and these will typically be estimated from a small amount of pilot data. Thus a parsimonious method of modelling correlations is needed if the simulation model is to be trained using pilot data. This issue is addressed here by assuming the $p$ variables are driven by a shared response to a few ($q \ll p$) unobserved latent variables through the use of factor analysis. These latent variables can be interpreted as unobserved environmental covariates \citep{warton2015so}. This approach requires $p(q+1) - q(q-1)/2$ elements to be estimated to formulate a covariance matrix $\Sigma_{FA}$, which can be much smaller than $p(p-1)/2$. For our fish abundance data set with $p=34$ species, if we take $1,2$ or $3$ factors we have  $68$, $101$ or $133$ covariance parameters to estimate, which is much less than the $561$ parameters that would otherwise have been needed.\\

This process has been implemented within the \texttt{ecopower} package, using an internal function called \texttt{extend}. This function takes a \texttt{cord} object  \citep[obtained by fitting a Gaussian copula to a \texttt{manyglm} object using the \texttt{cord} function from the \texttt{ecocopula} package;][]{popovic2021fast} and simulates \texttt{N} multivariate abundances using the above procedure (to date, it can handle Poisson, negative binomial and binomial distributions). The function then refits the simulated responses to a \texttt{manyglm} object with a data frame that is 'extended' in a manner that preserves the original or pre-specified design (Box 2). 

\subsection*{Challenge 2 -- Specifying an interpretable measure of effect size}
\label{ref:specifying parameters}

With a large number of taxa, there are a large number of ways that these taxa can respond to a treatment. In order to undergo a power analysis, an effect needs to be specified in a way that is interpretable, such that the researcher can understand how large the effect is in the context of their study, thus helping to understand whether or not it is necessary to increase the sample size. This interpretable measure of effect size also needs to be specified with relatively little \textit{a priori} information, which is generally lacking in an ecological setting (at least in comparison to the possible complex relations that the taxa can respond to a treatment).

A simple approach proposed here is to decide on:  
\begin{enumerate}
    \item Which species/response variables are expected to be (i) positively related to a given treatment (e.g. species that increase in abundance;``increasers"), (ii) those expected to be negatively related (species that decrease in abundance; ``decreasers"), or (iii) those not related to the treatment at all.
    \item The size of effect ($\rho$) that is negatively or positively related to the mean abundance $\mu$ of taxa, on the proportional scale. That is, $\rho=2$ means that the mean abundance doubles for increasers, and halves for decreasers. \\
\end{enumerate}

\noindent This is a relatively simplistic scenario, however it enables the effect size to be captured in a single coefficient $\rho$, and for expert opinion to inform the way in which different taxa are likely affected. Being on the proportional scale it also allows regression coefficients for simulated models to be easily specified, for example with log link; $\log{\rho}=\log{\dfrac{\hat{\mu}_{Treat=1}}{\hat{\mu}_{Cont=0}}}=\beta_{1j}$ for species that increase in abundance and $\log{1/\rho}=-\log{\dfrac{\hat{\mu}_{Treat=1}}{\hat{\mu}_{Cont=0}}}=-\beta_{1j}$ for species that decrease in abundance in the treatment group, relative to the control.\\

\begin{figure}[H]
					\centering
			\includegraphics[width=0.8\textwidth]{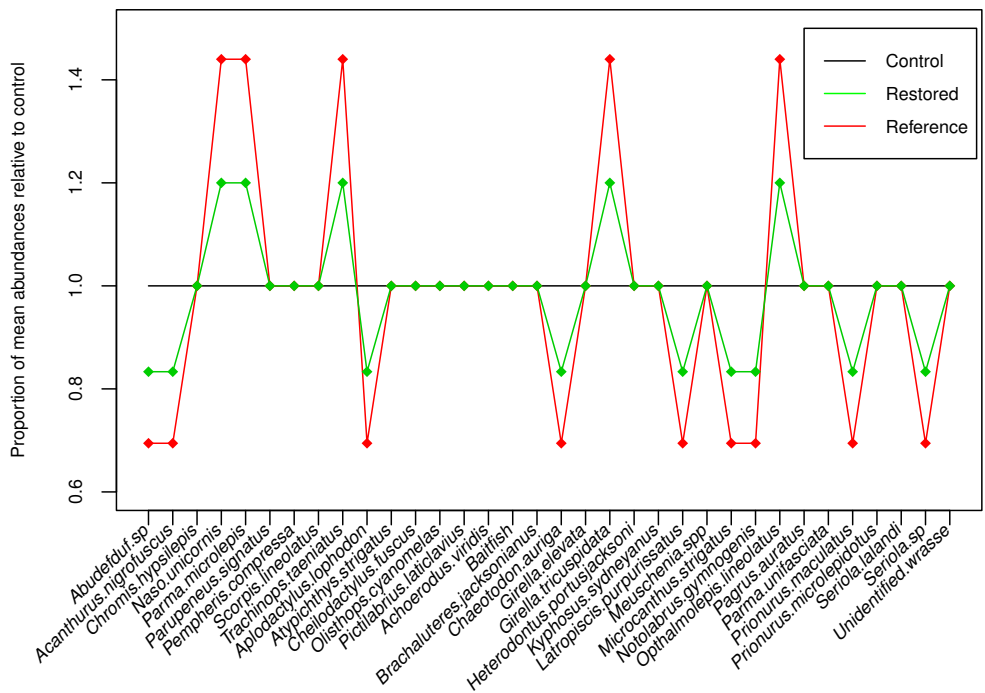}
			\caption{\small{Plot of mean abundance proportions in restored and reference sites relative to the control sites for a specified effect size of $\rho=1.2$.}}
			\label{fig:effectsizespecification}
	\end{figure}

For the Crayweed Restoration Project, we believe that restored sites will lie somewhere between reference and control sites \citep[e.g. because the extent of crayweed restored is still small relative to extant, natural populations;][]{layton2020kelp}, and by using existing data of fish surveys along the NSW coastline, as well as results from \citet{curley2002spatial}, assumptions were made as to which fish species will be expected to change in sites with crayweed (reference and restored), compared to those without crayweed (control). This allows the desired parameterisation of three types of fish species, whose mean abundances increase, decrease or do not change in reference and restored sites relative to control sites (Figure \ref{fig:effectsizespecification}, with the magnitude of these effects being specified with $\rho$ for changes between restored and control sites, and $\rho^2$ for changes between reference and control sites). \\

By using pilot data to estimate mean abundance of each species, then taking for example $\rho = 1.2$ to specify effects across different treatments, the mean abundances of data to be simulated from marginal distributions $F_j$ can be specified as in Figure \ref{fig:effectsizespecification}. $\rho-1$ can also be interpreted as the $\%$ change in mean abundance across treatments, so to answer our first experimental design question, we simply specify $\rho=1.2$.\\

This approach has been implemented within the \texttt{ecopower} package, with the \texttt{effect\_alt} function. Users input a \texttt{manyglm} object, the name of the predictor of interest (\texttt{term}), an effect size of interest (\texttt{effect\_size}) and a list of taxa that are ``\texttt{increasers}" or ``\texttt{decreasers}". The function then returns a parameterised coefficient matrix that can be used in ensuing power simulations, where taxa not specified as ``\texttt{increasers}" or ``\texttt{decreasers}" are assumed to not be affected. There are also options to specify more complicated effect sizes for effects that change over multiple levels of a categorical predictor. To produce the coefficient matrix in Figure \ref{fig:effectsizespecification} we would use the code in Box 1.

\begin{tcolorbox}
\begin{verbatim}
> library(ecopower)
> fit <- manyglm(fish~ Site.Type, family="negative.binomial",data = X)
> increasers <- c("Aplodactylus.lophodon","Atypichthys.strigatus",
+                 "Cheilodactylus.fuscus","Olisthops.cyanomelas",
+                 "Pictilabrius.laticlavius" )
> decreasers <- c("Abudefduf.sp","Acanthurus.nigrofuscus","Chromis.hypsilepis",
+                 "Naso.unicornis","Parma.microlepis","Parupeneus.signatus",
+                 "Pempheris.compressa","Scorpis.lineolatus","Trachinops.taeniatus")
> coeff.alt <- effect_alt(fit,effect_size=1.2,increasers,decreasers,
+                 term="Site.Type")    
\end{verbatim}
\end{tcolorbox}
Box 1: Code used to generate the effect size specification in Figure \ref{fig:effectsizespecification}.

%FIGURE 2

\subsection*{Challenge 3 -- Managing computation time}
\label{ref:managecomptime}

Power at significance level $\alpha$ can be estimated from a set of $n_{power}$ simulated datasets, where the $i$th dataset returned  $P$-value $p_i$, as follows:
	\[
			 \widehat{\text{Power}} = \dfrac{1}{n_{power}} \sum\limits^{n_{power}}_{i=1} I_{\{p_i\leq\alpha\}},
	\]

where $I_{\{\cdot\}}$ is the indicator function. That is, we estimate the power at significance level $\alpha$ as the sample proportion of $P$-values $p_i$ less than or equal to $\alpha$. In order to get a reliable Monte Carlo estimate of power we would conventionally use approximately $n_{power}= 1000$ datasets. In a multivariate abundance setting, the significance of multivariate Wald or score GEE test statistics is calculated via resampling techniques \citep{wang2012mvabund}. The standard asymptotic chi-squared techniques we would otherwise use are not suitable here because they assume large sample size ($n$) and a fixed number of responses ($p$), however for multivariate abundances, $p$ is rarely small compared to $n$.  Nevertheless, if resampling is used for testing, then for each $n_{power}$ simulated data set we would have to resample an additional $n_{resamp}\approx1000$ times to estimate a simulated p-value $p_i$ as $\hat{p}_i$. Each of these test statistics involve fitting a model to $p$ correlated taxa, which will involve fitting $p\times(n_{power}+n_{power}\times n_{resamp})$ GLMs overall and takes approximately 1.3 hours for a small data set of size $n=21$ and $p=34$ taxa, using parallel computing on a computer with 12 logical processes.\\% Others have had this issue in the past and have suggested doing...

A key innovation proposed in this paper is to use a critical value approach to estimate power, globally testing the significance of $n_{power}$ simulated test statistics $T_{1_i}$ using a critical value $\hat{c}_{\alpha}$. The critical value can be estimated as the upper $1 - \alpha$ qauntile of simulated test statistics under the null hypothesis $T_{0_j} \in H_0$ satisfying 
	 
	 \[
	 \dfrac{1}{n_{resamp}} \sum \limits^{n_{resamp}}_{j=1} \PP(T_{0_j}>\hat{c}_{\alpha}) = \alpha,
	 \]

and power can then be estimated as 

    \[
	 \widehat{\text{Power}}_{crit} = \dfrac{1}{n_{power}}\sum \limits^{n_{power}}_{i=1} I_{\{T_{1_i} > \hat{c}_\alpha\}}.
	 \]
In this approach power is estimated using test statistics $T_{1i}$ from simulated data, rather than $P$-values. This means that there is no longer a need to resample the simulated datasets, leading to a huge computational saving. The above approach involves only $n_{power}+n_{resamp} \approx 2000$ simulated or resampled datasets, a reduction by a factor of 500, and reducing computation time for the Crayweed restoration dataset from over an hour to just 42 seconds. Note this is an approximation to power, essentially, because we are assuming that $\hat{c}_{\alpha}$ is constant across all simulated datasets, when it may vary slightly. As sample size increases the accuracy of this approximation will improve, with the simulations below investigating how well this approximation works in our context.\\

The critical value approach to estimating power is implemented in the \texttt{ecopower} package, through the \texttt{powersim} function (which calls the internal \texttt{extend} function). It takes a \texttt{cord} object, a coefficient matrix (\texttt{coeffs}) that can be specified using the \texttt{effect\_alt} function, a  total sample size \texttt{N}, the name of the predictor of interest (\texttt{term}), number of simulations (\texttt{nsim}) and returns a power estimate. Computation time is also reduced by running simulations in parallel over a series of clusters (\texttt{ncores}), which defaults to one less then the number of cores available on your machine. To estimate power, for a sample size of $N=100$, using the pre-specified effect of $\rho = 1.2$ from \texttt{coeff.alt}, we would use the code in Box 2 (computation time is in seconds).

\begin{tcolorbox}
\begin{verbatim}
> fit_factors.cord = cord(fit)
> powersim(fit_factors.cord,N=100,coeff.alt,term="Site.Type",nsim=1000)
    Power Comp time 
  0.43700  42.02651 
\end{verbatim}
\end{tcolorbox}
Box 2: \texttt{ecopower} code to obtain a power estimate over \texttt{nsim=1000} simulations for a sample size of \texttt{N=100} using the \texttt{powersim} function by first fitting a copula model from our \texttt{manyglm} object using the \texttt{cord} function and then using the effect size (\texttt{coeff.alt}) generated in Box 1.

\section*{Results}

\subsection*{Validating method}
\label{ref:valmethod}

The effectiveness of estimating $\widehat{\text{Power}}$ as $\widehat{\text{Power}}_{crit}$ can be observed through direct application with simulations. Both methods are applied to answer our first experimental design question. That is, to find the number of samples required in order to likely detect an effect size of $20\%$ change in mean abundance of fish species between the three treatments (where some species have been specified to increase, decrease or not change across the treatments, Figure \ref{fig:effectsizespecification}).\\

\begin{figure}[H]
		\centering
		\includegraphics[width=.85\linewidth]{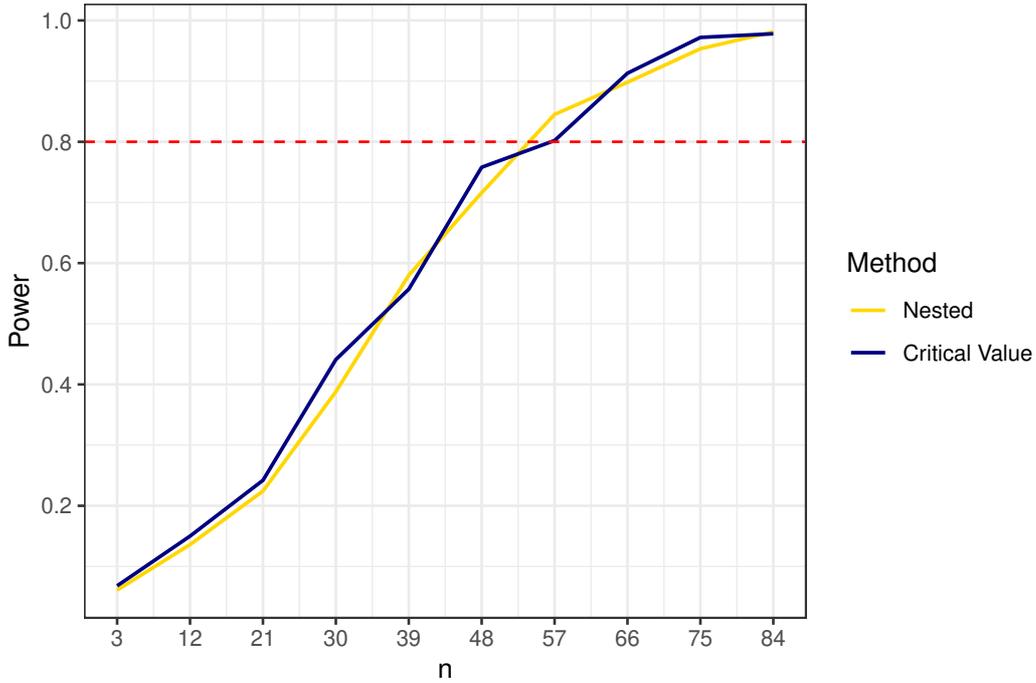}
		\caption{\small{In blue; power curve utilising a critical value approach $\widehat{\text{Power}}_{crit}$ (taking 11 minutes to compute with $n_{resamp}+n_{power}=2000$ simulated models). In gold; power curve utilising $n_{resamp}=1000$ and $n_{power}=1000$ simulated models with a nested approach to estimating power $\widehat{\text{Power}}$ (taking over 3 days to compute with $(n_{resamp}+1)n_{power}=1,001,000$ simulated models). Power has been estimated for a fish abundance data set ($p=34$ species) and an effect size specification of $20\%$ changes between three treatments. `n' is the number of samples per group, not the total sample size.}}
		\label{fig:simplot}
	\end{figure}

As can be seen in Figure \ref{fig:simplot}, the use of critical test statistics approximates power quite well, even for small study designs. Both approaches recommended at least $n^*=57$ sites per treatment group in order to likely detect $20\%$ changes in mean abundances. The critical value approach also reduced computation time from over 3 days to just 11 minutes. It tended to give slightly more variable power estimates, which could be overcome by increasing the number of resamples ($n_{resamp}$) or simulations ($n_{power}$) at the cost of increased computation time.  The critical value approach would still however be appreciably faster (unless $n_{resamp}$ and $n_{power}$ were increased to $500,000$!).\\

%FIGURE 3

\subsection*{Example}
\label{ref: Example}

While Figure \ref{fig:simplot} suggests $n^*=57$ sites per treatment group, we note this is not feasible under a balanced experimental design, because only 24 control or restored sites are available within the region where this restoration project was undertaken. Hence, it is unlikely that we will detect $20 \%$ changes in the mean abundances of fish species between the treatment levels. This is most likely due to the variability of the observed fish abundances creating a lot of noise in the data and the large ($66\%$) proportion of zero counts observed in the data.\\

\begin{figure}[H]
						\centering
			\includegraphics[width=.85\linewidth]{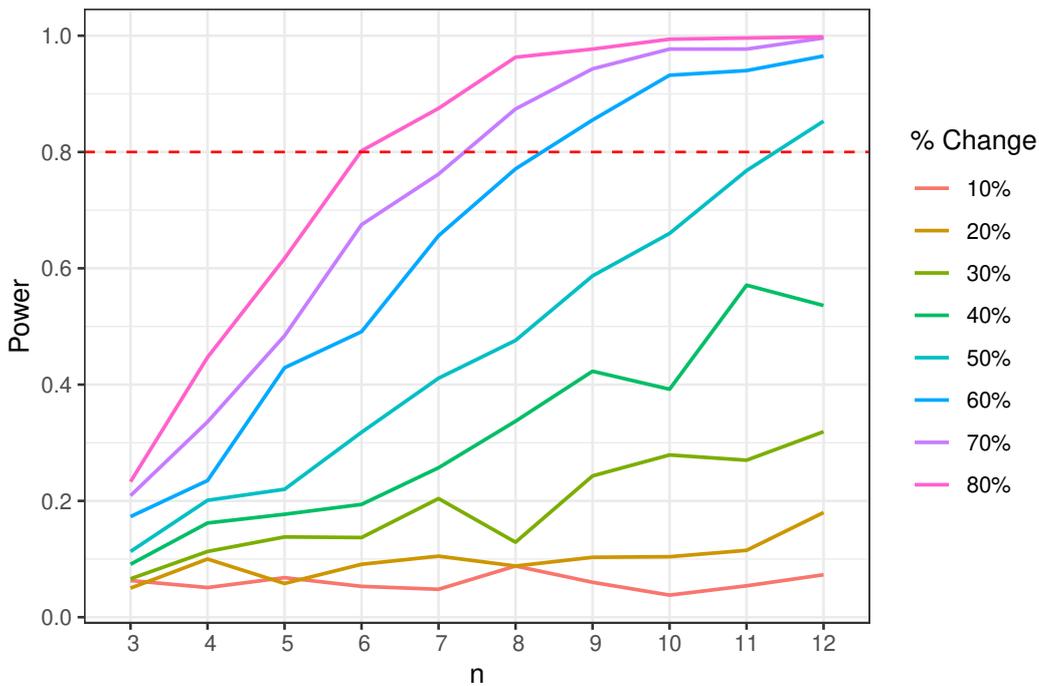}
			\caption{\small{Power curves for a range of effect size specifications $\rho_{10\%}=1.1,\rho_{20\%}=1.2,\rho_{30\%}=1.3,\rho_{40\%}=1.4,\rho_{50\%}=1.5,\rho_{60\%}=1.6,\rho_{70\%}=1.7,\rho_{80\%}=1.8$. 'n' is the number of samples per group, not the total sample size.}}
			\label{fig:powereffcurve.manyglm}			
		\end{figure}

	The second experimental design question we are interested in is: what effect sizes are likely to be observed at 12 sites per treatment? In order to answer this question, we can simply plot power curves over a range of effect size specifications. We actually plotted power curves as a function of sample size $n$ for a broader perspective (Figure \ref{fig:powereffcurve.manyglm}), for effect size specifications $\rho_{10\%}=1.1,\rho_{20\%}=1.2,\rho_{30\%}=1.3,\rho_{40\%}=1.4,\rho_{50\%}=1.5,\rho_{60\%}=1.6,\rho_{70\%}=1.7,\rho_{80\%}=1.8$. Computing each power curve took approximately 5 minutes of computation time on a computer with 12 logical processors, less time than for Figure \ref{fig:simplot}, because the sample sizes under consideration here were smaller. We observe as expected that as we increase the effect size specifications, we increase the power of each experimental design (Figure \ref{fig:powereffcurve.manyglm}), since larger effect sizes are easier to detect. Importantly, we observe that under the maximum balanced sample size of $n^*=12$ sites per treatment, the smallest effect size likely to be observed is $\rho_{50\%}=1.5$, or equivalently $50 \%$ changes in the mean abundances of fish species. The pilot survey included $n=6$ sites per treatment, which would only be able to detect quite extreme effects.

%FIGURE 4

	\section*{DISCUSSION}
	
	Here we have described a sample size estimation procedure for multivariate abundance data. Techniques and software for this are much needed in ecology, however the problem is rather difficult technically, given the three aforementioned challenges. The main innovations of our procedure are to: specify a joint data generating model through discrete margin copulas that is able to be tuned to the properties of the pilot data; implement a simple and interpretable approach to specifying an effect size for complex multivariate effects; reduce computation time by a factor of 500, which was achieved using a ``critical value" approach to power estimation. The procedure has been coded in \texttt{R} with general purpose functions in the \texttt{ecopower} package, which can be downloaded from Cran \citep{maslen2021package}. This procedure can be applied to a vast range of ecological studies with multivariate abundance data to answer experimental design questions and give sample size recommendations, an important yet technically difficult task not previously addressed in the literature.\\
	
    Being a general-purpose procedure, power can also be estimated for more complex designs. For instance, the methods proposed here (and the software tools in the \texttt{ecopower} package) can estimate power for effects on both categorical and quantitative predictors, handle the most common distributions for handling abundance or presence-absence data (to date; Poisson, negative binomial and binomial distributions) and handle designs with multiple covariates. It can also investigate unbalanced designs. In our fish abundance example, we could study the change in power when fixing the number of control and restored sites at $n_1=n_2=12$, but varying the number of reference sites, as more of these are available regionally. It is also possible to use \texttt{ecopower} to specify different and potentially more complicated effect size structures, if desired. Note however that all effect size parameters need to be set \textit{a priori}, and the more intricate the scenario, the more \textit{a priori} decisions would be needed.\\
	
   It would be interesting to apply this procedure to other study designs and consider whether the patterns seen here apply elsewhere too. Specifically, this study's intended sampling design ($n=6$) was underpowered for the size of effects that we would like to detect. Is this more generally true -- do ecological monitoring studies tend not to sample enough sites to detect effects of practical interest?

\subsection*{Acknowledgements}
David Warton and Gordana Popovic are supported by the Australian Research Council Discovery Grant DP210101923. Ezequiel Marzinelli and Adriana Verges were both supported by the Australian Research Council Linkage Grant LP160100836, the NSW DPI Recreational Fishing Trust, and Environmental Trust. Ezequiel Marzinelli and Adriana Verges were also supported by Australian Research Council Discovery Grants DP180104041 and DP190102030 respectively. We would also like to thank Alexandra Campbell for field sampling.

\subsection*{Conflict of interest}
The authors declare no conflict of interest.

\subsection*{Author contributions}
Ben Maslen designed the power analysis procedure and built the R package ecopower. David Warton and Gordana Popovic provided expertise and direction in copula modelling, multivariate analysis and power simulation. Adriana Verges and Ezequiel Marzinelli provided the fish abundance data, the motivating example and used their domain expertise to help define the effect size of interest. Ben Maslen wrote the first draft of the manuscript, and all authors contributed substantially to revisions.

%\subsection*{Data availability} 
%Should the manuscript be accepted, the data will be archived in the Dryad public repository and the data DOI will be included at the end of the article.

%	\bibliographystyle{abbrv}
%	\bibliographystyle{agsm}
	\bibliographystyle{apalike}
	\bibliography{Main_Document}
	
%\section*{Figures}

\section*{Supporting information}

\end{document}